\documentclass[twoside]{dis07}
\usepackage[latin1]{inputenc}
\usepackage[dvips]{graphicx,epsfig,color}
\usepackage{wrapfig,rotating}
\usepackage{amssymb,amsmath,array}

\newcommand{\xbj}{{x}}

\newcommand{\qssa}{{Q^2_\mathrm{s,A}}}

\newcommand{\qssb}{{Q^2_\mathrm{s,B}}}

\newcommand{\qssp}{{Q^2_\mathrm{s,p}}}
\newcommand{\qs}{{Q_\mathrm{s}}}
\newcommand{\qss}{{Q^2_\mathrm{s}}}

\newcommand{\as}{{\alpha_{\mathrm{s}}}}
\newcommand{\rt}{{\mathbf{r}_\perp}}

\newcommand{\bt}{{\mathbf{b}_\perp}}

\newcommand{\nc}{{N_\mathrm{c}}}

\newcommand{\gev}{\textrm{ GeV}}
\newcommand{\fm}{\textrm{ fm}}
\newcommand{\mb}{\textrm{ mb}}
\newcommand{\ra}{R_A}
\newcommand{\rp}{R_p}

\newcommand{\Aavg}[1]{\left\langle #1 \right\rangle_\textrm{N}}

\newcommand{\nr}[1]{(\ref{#1})} 
\newcommand{\ud}{\, \mathrm{d}}

\newcommand{\eq}{Eq.~}

\newcommand{\sigmap}{{ \sigma^\textrm{p}_\textrm{dip} }}

\newcommand{\dsigmap}{{\frac{\ud \sigma^\textrm{p}_\textrm{dip}}{\ud^2 \bt}}}
\newcommand{\dsigmaa}{{\frac{\ud \sigma^A_\textrm{dip}}{\ud^2 \bt}}}

\newcommand{\dsigma}{{\frac{\ud \sigma_\textrm{dip}}{\ud^2 \bt}}}

\begin{document}
\title{Universal features of QCD dynamics\\ in hadrons and nuclei at high energies}

\author{Raju Venugopalan
%
\thanks{This work is supported by DOE Contract No. DE-AC02-98CH10886. I thank T. Lappi and C. Marquet 
for their comments on the manuscript.}
%
\vspace{.3cm}\\
%
Physics Department, Brookhaven National Laboratory \\
Upton, NY 11973, USA
%
}

\maketitle

\begin{abstract}

We discuss the empirical evidence for a universal Color Glass Condensate and outline prospects for further 
studies at future colliders. Some ramifications for initial conditions in heavy ion collisions are pointed out.

\end{abstract}

\section{Introduction}

QCD has been called the perfect theory~\cite{Wilczek}; as a renormalizable field theory whose validity 
could extend up to the grand unification scale, it provides the mechanism for generating nearly all the mass of the visible universe. The current quark masses are the only external parameters in the theory. Quenched QCD, without dynamical quarks, explains the hadron spectrum to an accuracy~\footnote{Some lattice QCD computations with dynamical quarks 
claim improved agreement to within a few percent~\cite{Lattice-review}.} of ~10\%. These lattice results suggest that gluons play a central role in the structure of matter. 

The role of glue in QCD is best understood in the asymptotic weak coupling regimes of the theory where 
analytical computations are feasible. Much of the discussion in perturbative QCD (pQCD) has been in the Bjorken-Feynman asymptotics where $Q^2 \longrightarrow \infty$, $s\longrightarrow \infty$ and 
$x_{\rm Bj} \equiv Q^2/s = {\rm fixed}$. The machinery of precision physics in QCD such as the operator product expansion 
and factorization theorems are derived in this limit of the theory. The progress in this direction has been truly remarkable~\cite{Vogt}. In DIS for instance, both coefficient functions and splitting functions have been derived to next-to-next-to leading order (NNLO).    

What does the hadron look like in the Bjorken-Feynman asymptotics? The DGLAP evolution equations tell us that the 
gluon distribution grows rapidly with increasing $Q^2$ at small $x_{\rm Bj}$. However, the phase space density (in a particular gauge and frame), decreases rapidly with increasing $Q^2$. The proton become more ``dilute" even though the number of partons increases; the typical size of resolved partons decreases as $1/Q^2$, 
faster than the increase in the number through QCD evolution. The more dilute the hadron, the cleaner will be the QCD background for new physics beyond the standard model. 

Much of the current focus in QCD is in quantifying this background. It would be unfortunate however if this were the only focus in QCD studies because the theory, even in the weak coupling domain, contains rich and non-trivial dynamics.  We speak here of the Regge-Gribov asymptotics where $x_{\rm Bj}\longrightarrow 0$, $s\longrightarrow \infty$ and $Q^2 = {\rm fixed}$. This regime of strong color fields is responsible for the bulk of multiparticle production in QCD. 
What does the hadron look like in the Regge-Gribov asymptotics ? The BFKL equation, which resums the leading logarithms in $x$, indicates that the gluon distributions grow even more rapidly in this asymptotics. Unlike the Bjorken-Feynman case, the phase space density in the hadron grows rapidly as well. The stability of the theory requires that the phase space densities (or more generally, the field strengths squared) be no larger than $\sim 1/\alpha_S$. 
In the pQCD framework, mechanisms for the saturation of the growth in the phase space density are provided by ``higher twist" recombination and screening contributions~\cite{GLRMQ}. These counter the bremsstrahlung growth of soft gluons described by the DGLAP and BFKL equations. The saturation scale $Q_s(x)$  generated by the dynamics demarcates the 
separation between the linear and non-linear regimes of the theory: for momenta $Q^2 \ll Q_s^2$, non-linear QCD dynamics is dominant, for momenta $Q^2 \gg Q_s^2$, weak coupling physics is governed by the DGLAP/BFKL evolution equations. 

The universal properties of gluons in the non-linear regime are described by a classical effective field theory of dynamical 
gluon fields coupled to static, stochastic sources. This is the Color Glass Condensate (CGC)~\cite{CGC}. The evolution of 
multi-parton correlators  with energy is described by the Wilsonian JIMWLK renormalization group (RG) equations~\cite{JIMWLK}. In the limit of large nuclei and large $N_c$, one recovers the Balitsky-Kovchegov equation for the forward 
dipole cross-section~\cite{BK}. A universal saturation scale arises naturally in the theory and its energy dependence is given by the JIMWLK/BK equations. The typical momentum of gluons $\sim Q_s \gg \Lambda_{\rm QCD}$;  the bulk of the contributions to high energy cross-sections may be therefore described in a weak coupling framework.
 
 The saturation scale also grows with the nuclear size. A fast compact probe of size $1/Q_s < R_p$, where $R_p$ is the proton size, will interact coherently at high energies with partons localized in nucleons all along the nuclear diameter. The field strength squared experienced by the probe is therefore enhanced parametrically by a factor proportional to the nuclear diameter $\sim A^{1/3}$. As clearly illustrated in the CGC effective theory, the dynamics of partons at small $x$ is universal regardless of one speaks of hadrons or nuclei; the latter, as we will discuss further, are therefore an efficient (and cheaper) amplifier of the non-linear dynamics of these gluons.   
 
This talk is organized as follows. We will outline our current (limited) understanding of the different dynamical regimes in high energy QCD from experiments at HERA and RHIC. We will then discuss how experiments at the LHC and future DIS experiments on nuclei can help further quantify our understanding. Finally, to illustrate the scope of these studies, we will discuss how the strong color field dynamics of partons in nuclear wavefunctions contributes to a quantitative understanding of the formation and subsequent thermalization of a strongly interacting ``glasma" in heavy ion collisions. 

\section{The evidence for the CGC from e+p DIS}

\begin{figure}[htbp]
\begin{center}
\includegraphics[width=0.32\linewidth]{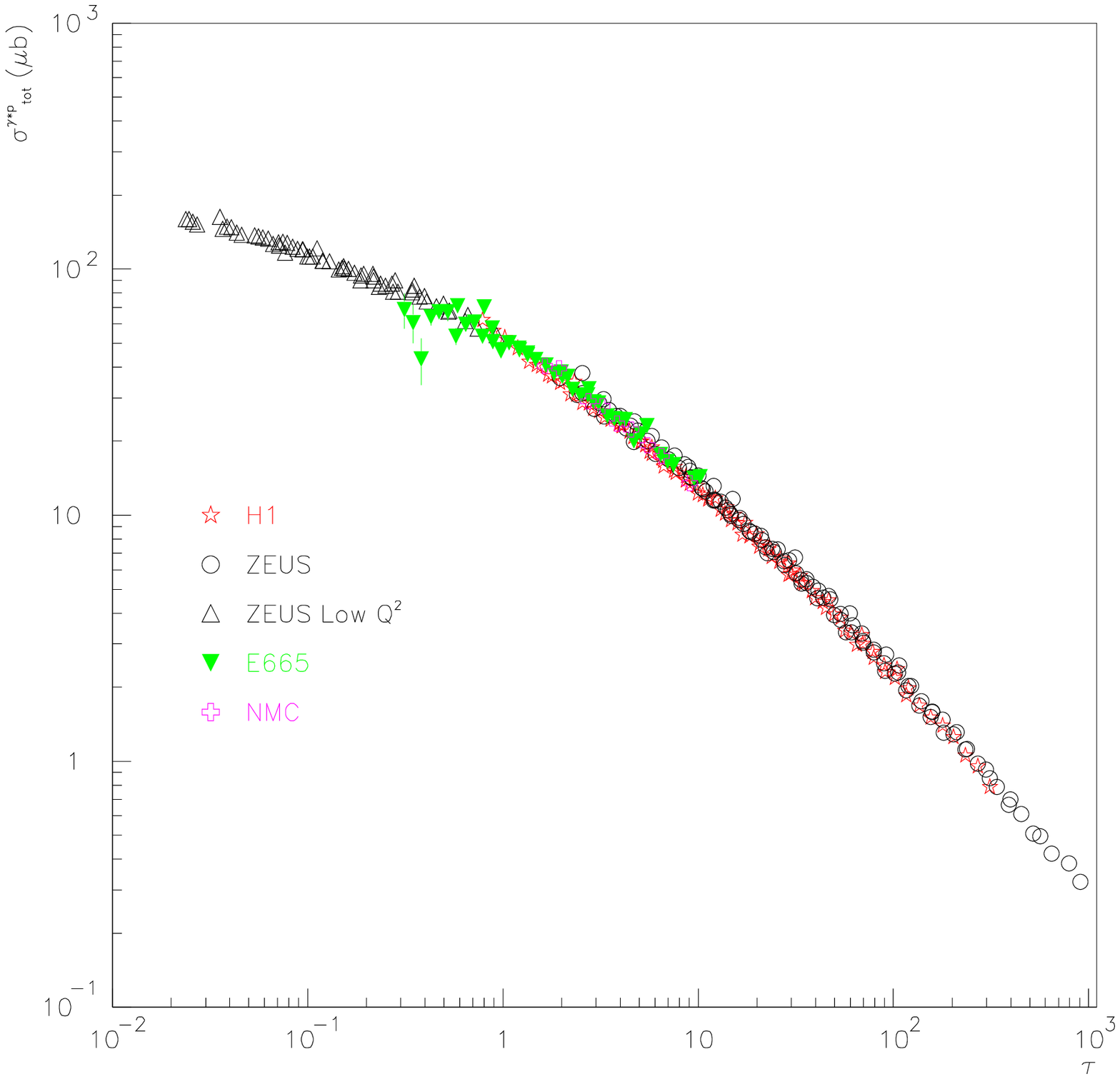}
\includegraphics[width=0.32\linewidth]{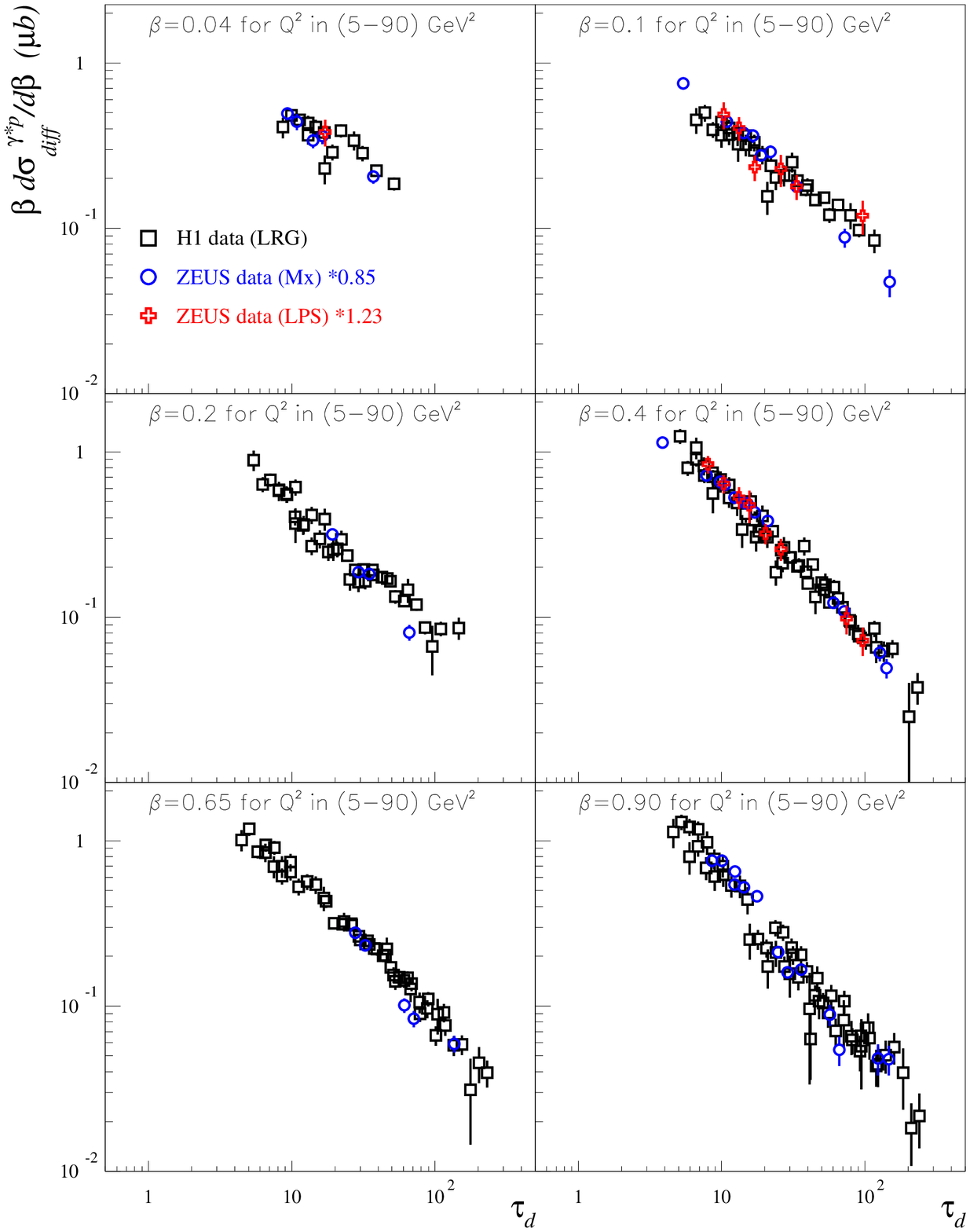}
\includegraphics[width=0.32\linewidth]{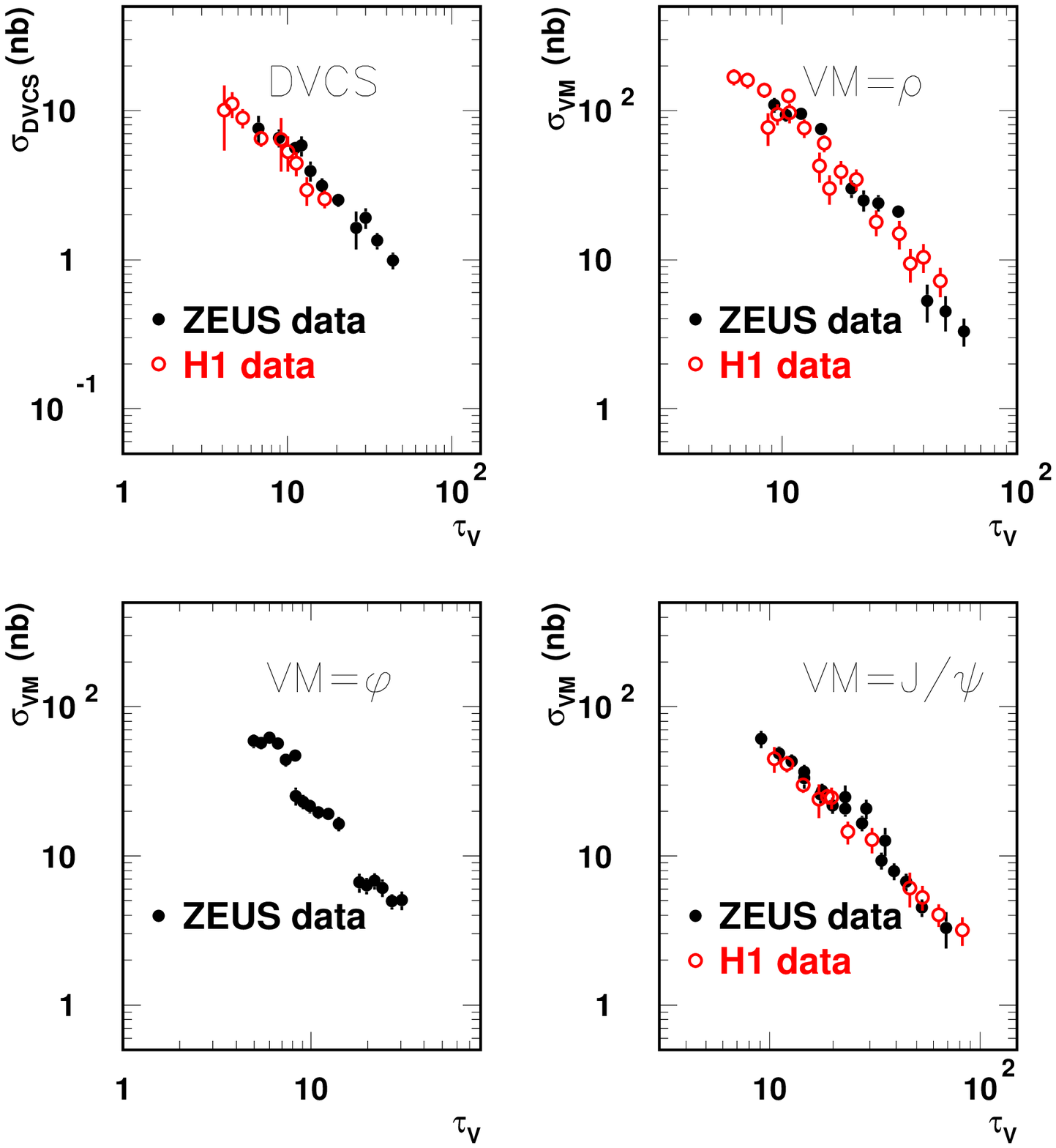}
\end{center}
\caption{Geometrical scaling fully inclusive, diffractive and exclusive vector meson cross-sections. From ~\cite{MarquetSchoeffel}.}
\label{fig:scaling}
\end{figure}

A strong hint that semi-hard scales may play a role in small $x$ dynamics at HERA came from ``geometrical scaling" of the HERA data~\cite{GolecSK}. The inclusive virtual photon+proton cross-section for $x\le 0.01$ and 
all available $Q^2$ scales~\footnote{The E665 data are a notable exception.} as a function of $\tau\equiv Q^2/Q_s^2$, where $Q_s^2(x) = \exp(\lambda Y)$ GeV$^2$. Here $Y=\ln(x_0/x)$ is the rapidity;  $x_0= 3\cdot 10^{-4}$ and $\lambda= 0.288$ are parameters fit to the data~\cite{GolecSK,MarquetSchoeffel}. Further, the inclusive diffractive, vector 
meson and DVCS cross-sections at HERA, with a slight modification~\footnote{$\tau_{D,VM} = (Q^2 + M^2)/Q_s^2$, where 
$M$ denotes the mass of the diffractive/vector meson final state.}in the definition of $\tau$, also appear to show geometrical scaling~\cite{MarquetSchoeffel}.  Geometrical scaling of the e+p data is shown in Fig.~\ref{fig:scaling}. A recent ``quality factor" statistical analysis~\cite{Gelis-etal} indicates that this scaling is robust; it is however unable to distinguish between the above fixed coupling energy dependence of $Q_s$ and the running coupling $Q_s(x)\propto \exp(\sqrt{Y})$ dependence of the saturation scale. 
Geometrical scaling is only asymptotic in both fixed and running coupling evolution equations~\footnote{The effect of 
``pomeron loops" on this scaling will be discussed later.}. 
Pre-asymptotic corrections have been computed previously, to good approximation, in both fixed and running 
coupling cases for the BK equation~\cite{MunierPeschanski}. A recent NLO BK analysis~\cite{AlbaceteKovchegov} suggests that the onset of the scaling asymptotics may be precocious~\footnote{For another very interesting take on precocious scaling, see Ref.~\cite{Cyrille}.}, thereby providing a possible explanation for its manifestation in the HERA data. A caveat that has been raised is that there is a strong correlation between $x$ and $Q^2$ in the HERA 
data~\cite{AvsarGustafson}. The scaling however persists even where there is a significant lever arm in $Q^2$ for small 
$x$. Nevertheless, geometrical scaling alone is not sufficient evidence of saturation effects and it is important to look at the 
data in greater detail in saturation/CGC models. 

All saturation models~\cite{Mueller:1989st} express the inclusive virtual photon+proton cross-section as 
\begin{equation}\label{eq:sigmatot}
\sigma^{\gamma^*p}_{L,T}
= \int\! \ud^2 \rt \int_0^1 \! \ud z \left| \Psi^{\gamma^*}_{L,T}
\right|^2 
\int \! \ud^2 \bt \dsigmap .
\end{equation}
Here $\left| \Psi_{L,T}^{\gamma^*}(\rt,z,Q) \right|^2$
represents the probability for a  virtual photon to produce a quark--anti-quark pair of size $r = |\rt|$ and $\dsigmap(\rt,\xbj,\bt)$ denotes the \emph{dipole cross section} for this pair to scatter off the target at an impact parameter $\bt$. The former is well known from QED, while the latter represents the dynamics of QCD scattering at small $x$. A simple saturation model 
(known as the GBW model~\cite{Golec-Biernat:1998js}) 
of the dipole cross section, parametrized as $\dsigmap = 2 ( 1 - e^{ - r^2 \qssp(x)/4})$ where 
$\qssp (x) = (x_0/x)^\lambda \gev^2$, gives a good qualitative fit to the HERA inclusive cross section data for 
$x_0 = 3\cdot 10^{-4}$ and $\lambda = 0.288$. Though this model captures the qualitative features of saturation, it does not contain the bremsstrahlung limit of perturbative QCD (pQCD) that applies to small dipoles of size $r \ll 1/\qs(x)$. 

In the classical effective theory of the 
CGC, one can derive, to leading logarithmic accuracy, the dipole cross section~\cite{Venugopalan:1999wu} 
containing the right small $r$ limit. This dipole cross section can be 
represented  as~\cite{Kowalski:2003hm}
\begin{equation}
\dsigmap
 = 2\,\left[ 1 - \exp\left(- r^2  F(\xbj,r) T_p(\bt)\right) 
\right],
\label{eq:BEKW}
\end{equation}
where $T_p(\bt)$ is the impact parameter profile function in the proton, normalized as 
$\int d^2 \bt \,T_p(\bt) = 1$ and $F$ is proportional to the gluon distribution~\cite{Bartels:2002cj}
\begin{equation}
F(\xbj,r^2) = \pi^2 \as\left(\mu_0^2 + 4/r^2 \right) 
\xbj g\left(\xbj,\mu_0^2 + 4/r^2 \right)/(2 \nc)\,,
\label{eq:BEKW_F}
\end{equation}
evolved from the initial scale $\mu_0$ by the DGLAP equations. The dipole cross section in \eq\nr{eq:BEKW} was 
implemented in the impact parameter saturation model (IPsat)~\cite{Kowalski:2003hm} where the parameters are fit to 
reproduce the HERA data on the inclusive structure function $F_2$. Here $\qs$ is defined as the solution 
of $\dsigma(\xbj , r^2 = 1/\qss(\xbj,\bt)) = 2(1-e^{-1/4})$~\footnote{
This choice of  is equivalent to the 
saturation scale in the GBW model for the case of a Gaussian dipole cross section.}.

The IPsat dipole cross section in \eq\nr{eq:BEKW} is valid when leading logarithms in 
$x$ in pQCD are not dominant over leading logs in $Q^2$. At very small $x$, where 
logs in $x$ dominate, quantum evolution in the CGC describes both the BFKL limit of linear small $x$ evolution as well as nonlinear JIMWLK/BK evolution at high parton densities~\cite{JIMWLK,BK}.
These asymptotics are combined with a more realistic $b$-dependence
in the b-CGC model~\cite{Iancu:2003ge,Kowalski:2006hc}. Both the IPsat model and 
the b-CGC model provide excellent fits to HERA data for $x \leq 0.01$~\cite{Kowalski:2006hc,Forshaw:2006np}. 
\begin{figure}[htbp]
\begin{center}
\resizebox*{6cm}{!}{\includegraphics{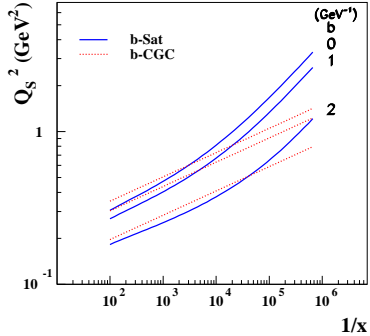}}
\end{center}
\caption{\label{fig:Qs} The saturation scale vs $1/x$ in the IPsat and b-CGC models~\cite{Kowalski:2006hc}.}
\end{figure} 
An important caveat~\cite{Thorne}  to the success of the saturation models is that the saturation scale, at median impact parameters, extracted from these fits is $\leq 1$ GeV$^2$ even at the lowest $x$ values at HERA~\cite{Kowalski:2006hc,Golec-Sapeta}. The saturation scale extracted from the fit in the IPsat model is shown in Fig.~\ref{fig:Qs}. We should note however that the uncertainty in the magnitude of the saturation scale is significant and is a factor of 2 larger in recent 
CGC fits~\cite{Soyez-07}. NLO computations in the small $x$ dipole framework are now becoming available~\cite{FadinBWK}; these will provide theoretical guidance into precisely how the coupling runs as 
a function of $\qs$ at small $x$. Finally, from Fig.~\ref{fig:Qs}, it is clear that the energy dependence of the extracted 
$\qs$ is significantly stronger than those predicted in non-perturbative models~\cite{DonnachieLandshoff}.

\section{The evidence for the CGC from e+A DIS and d+A and A+A collisions}
\begin{figure}[htbp]
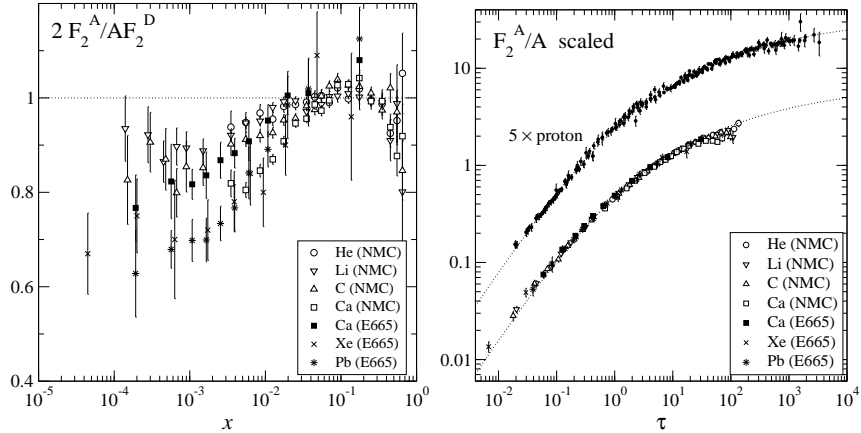

\begin{center}
\includegraphics[width=0.4\linewidth]{ratio_all_notscaled_bw.eps}
\includegraphics[width=0.4\linewidth]{F2_N+P_scaled_bw.eps}
\end{center}
\caption{Left: Shadowing of $F_2$ from the NMC and E665 fixed target experiments. Right: The data scaled as a 
function of $\tau\equiv Q^2/Q_s^2$~\cite{Freund:2002ux}.}
\label{fig:shad}
\end{figure}

\begin{figure}[htbp]
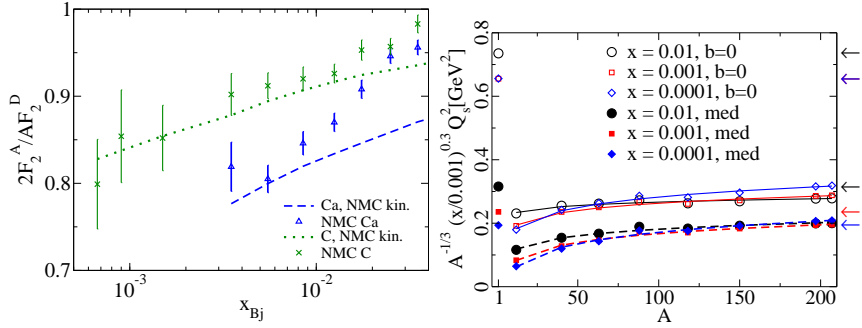

\begin{center}
\includegraphics[width=.4\linewidth]{glaubercompilationnmc.eps}
\includegraphics[width=.4\linewidth]{Qscombination.eps}
\end{center}
\caption{Left: Comparison of the IPsat model (with no adjustable parameters) to the NMC data. Right:  The 
$A$ and $x$ dependence of the saturation scale in the IPsat model~\cite{HTR}.}
\label{fig:shad2}
\end{figure}

The strong field dynamics of small $x$ partons is universal and  should be manifest in 
large nuclei at lower energies than in the proton. In Fig.~\ref{fig:shad} (left), we show 
the well known shadowing of $F_2^A$ in the fixed target e+A E665 and NMC experiments. Expressed in terms of 
$\tau\equiv Q^2/Q_s^2$ (Fig.~\ref{fig:shad} (right)), the data show geometrical scaling~\cite{Freund:2002ux}. In Ref.~\cite{Freund:2002ux}, the $A$ dependence of $Q_s$ is determined to be $A^{1/4}$ and not $A^{1/3}$ as suggested in a simple random walk picture. However, as we shall discuss shortly, this conclusion is a little misleading. 
A study of nuclear DIS in the IPsat CGC framework was performed in Ref.~\cite{Kowalski:2003hm,HTR}. The 
average differential dipole cross section is well approximated by $\Aavg{\dsigmaa} 
\approx 2\left[1-\left(1-\frac{T_A(\bt)}{2}\sigmap \right)^A\right]$, where $T_A(\bt)$ is the well known Woods Saxon distribution. Here $\sigmap$ is determined from the IPsat fits to the e+p data; no additional parameters are introduced for $eA$ collisions. In Fig.~\ref{fig:shad2} (left), the model is compared to 
NMC data on Carbon and Calcium nuclei-the agreement is quite good. In Fig.~\ref{fig:shad2} (right), we show the extracted 
saturation scale in nuclei for both central and median impact parameters. To a good approximation~\footnote{This is  considerably larger than the simplest estimate of a $\theta$-function impact parameter dependence in the GBW model, which yields $\qssa \approx A^{1/3}\frac{\rp^2 A^{2/3}}{\ra^2}\qssp\approx 0.26 A^{1/3} \qssp$ for $ 2\pi \rp^2 \approx 20\mb $ and $\ra\approx 1.1\,A^{1/3}\fm$.},  the saturation scale in nuclei scales as $\qssa(x,b_{\rm med.}) \approx \qssp(x,b_{\rm med.})\cdot (A/x)^{1/3}$.  The factor of $200^{1/3} \approx 6$ gives a huge ``oomph'' in the parton density of a nucleus relative to that of a proton at the same $x$. \emph{Indeed, one would require a center of mass energy 
$\sim 14$ times larger~\footnote{At extremely high energies, this statement must be qualified to account for the 
effects of QCD evolution~\cite{Mueller:2003bz}.} 
in an e+p collider relative to an e+Au collider to obtain the same $\qssa(b_\textrm{med.})$.} 
The reasons for the additional enhancement are two fold. Firstly, because the density 
profile in a nucleus is more uniform than that of the proton, $\qssp(b_\textrm{med.})$ is  only $\sim 35$\% of the value at 
$b=0$; in contrast, in gold nuclei it is 70\%. Because the median impact parameter dominates inclusive scattering, this 
effect gives a significant enhancement to the effective $Q_s$. The second reason for the enhancement is the DGLAP-like growth of the gluon distribution in the IPsat nuclear dipole cross section. For two  nuclei, $A$ and $B$ (with $A>B$), in a ``smooth nucleus'' approximation ($\sum_{i=1}^A T_p(\bt-\bt_i) \longrightarrow A\,T_A(\bt)$), 
$\frac{\qssa}{\qssb} \approx  \frac{A^{1/3}}{B^{1/3}} \frac{F(\xbj,\qssa)}{F(\xbj,\qssb)}$, where $F$ was defined in eq.~\ref{eq:BEKW_F}. The scaling violations in $F$ imply that, as observed in Refs.~\cite{Kowalski:2003hm,Armesto:2004ud},
the growth of $\qs$ is faster than $A^{1/3}$. Also, because the increase  of $F$ with $Q^2$ is faster for smaller $x$, the 
$A$-dependence of $\qs$ is stronger for higher energies. In contrast, the dipole cross section in the b-CGC model depends only on the combination~\footnote{With the caveat that it has BFKL-like 
violations that vanish asymptotically with $Y$.} $r \qs(x)$ without DGLAP scaling violations.  It therefore does not have this 
particular nuclear enhancement. Another interesting possibility, following from running 
coupling corrections to the leading logs in $x$, is that QCD evolution actually depletes the nuclear enhancement of $\qs$ at very small $x$~\cite{Mueller:2003bz}.  {\it Precise extraction of the $A$ dependence of $\qs$ can therefore help distinguish between ``classical'' and ``quantum'' RG evolution at small x}. 

We now turn to a discussion of CGC effects in hadronic collisions. A systematic 
treatment of the scattering of two strong color sources (such as two high energy nuclei) is discussed in Section~\ref{sec:Glasma}. To leading order, the problem reduces to the solution of the classical Yang-Mills (CYM) equations averaged over color sources for each nucleus~\cite{KovnerMW,KV}; the variance of this distribution of sources is proportional to $\qssa$. Besides the nuclear radius, $\qssa$ is the only scale in the problem, and 
the $\qssa \sim \qssp \cdot (A/x)^{0.3}$ expression for the saturation scale was used in CGC models of nuclear 
collisions to successfully predict the multiplicity~\cite{KV} and the centrality dependence of the multiplicity~\cite{KN} dependence in gold+gold collisions at RHIC. The universality of the saturation 
scale also has a bearing on the hydrodynamics of the Quark Gluon Plasma (QGP); the universal form leads to a lower 
eccentricity~\cite{LappiV} (and therefore lower viscosity) than a non-universal form that generates a larger eccentricity~\cite{Hirano-etal} (leaving room for a larger viscosity) of the QGP.

\begin{figure}[htbp]
\begin{center}
\resizebox*{6cm}{!}{\includegraphics{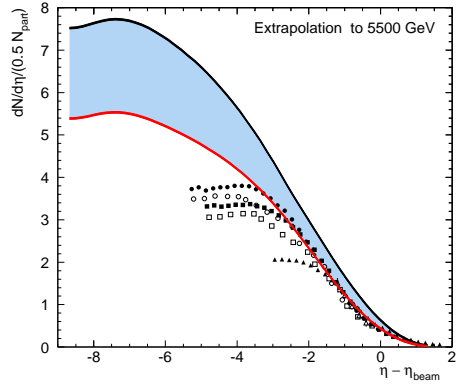}}
\end{center}
\caption{\label{fig:LF}Prediction for limiting fragmentation 
and deviations away from it at LHC energies. The bands denote the range in the predictions for GBW and MV models. From 
~\cite{FAR}. }
\end{figure}

\begin{figure}[htbp]
\begin{center}
\includegraphics[width=.4\linewidth,angle=270]{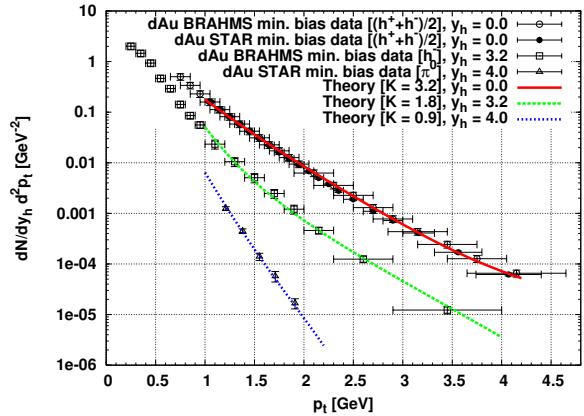}
\end{center}
\caption{The inclusive $k_\perp$ distributions in deuteron+gold collisions compared to theory curves for different 
rapidities. From ~\cite{DumitruHJ}.}
\label{fig:RdA}
\end{figure}

For asymmetric (off-central rapidity) nuclear collisions, or proton/deuteron+heavy nucleus collisions, $k_\perp$-factorization 
can be derived systematically for gluon production, at leading order, in the CGC framework~\cite{KovchegovMueller}. The simplicity of $k_\perp$ factorization is convenient for 
phenomenology; predictions based on this formalism describe the rapidity distributions in $A+A$ 
collisions~\cite{KL} and the phenomenon of ``limiting fragmentation"~\cite{Jalilian}. The latter, and deviations thereoff, are 
described by solutions of the BK-equation. Predictions for the multiplicity distribution in A+A collisions at the LHC~\cite{FAR} for both GBW  and classical CGC (MV) initial conditions~\footnote{The McLerran-Venugopalan 
(MV) initial condition has the same form as the IPsat dipole cross-section discussed earlier.} give a charged particle multiplicity of 1000-1400 in central lead+lead collisions at the LHC. The results are shown in Fig.~\ref{fig:LF}.
In deuteron+gold collisions at RHIC, the normalized ratio $R_{pA}$ of the inclusive hadron spectrum relative to the same in proton+proton collisions shows a mild ``Cronin" peak at mid-rapidities corresponding to multiple scattering in the 
classical CGC; at forward rapidities, however, $R_{pA}$ decreases systematically below unity. In the CGC, this reflects quantum evolution of the dipole cross-section in a 
large nucleus and has the same origin as the extension of the geometrical scaling regime~\cite{EKL} to $Q \gg Q_s$. This effect~\footnote{Quantum evolution here corresponds to the BK anomalous dimension of $\gamma = 0.63$ in the dipole cross-section, as opposed to $\gamma =1$ (DGLAP) and $\gamma=0.5$ (BFKL).},  should 
also exist in hadronic collisions~\cite{KharzeevLM}; specifically, it was predicted this would occur in deuteron+gold collisions~\cite{KKT}. In general, $R_{pA}$ while suggestive, is not an ideal variable because it is not clear the same formalism applies 
to p+p collisions at the same rapidity. Data on the inclusive hadron spectrum in deuteron+gold collisions can be directly compared to model predictions~\cite{DumitruHJ}~\footnote{The same analysis also gives good agreement for the forward 
p+p spectrum at RHIC~\cite{BoerDH}.}. The result is shown in Fig.~\ref{fig:RdA}. For a comprehensive review 
of applications of CGC picture to RHIC phenomenology, we refer the reader to Ref.~\cite{YJ2}. 
There are a couple of caveats to this picture. Firstly, $k_\perp$ factorization is very fragile. It does not hold for quark production even at leading order in the parton density~\cite{BlaizotGV2}, albeit it may be a good approximation for large masses and transverse momenta~\cite{GelisV1}. For gluon production, it does not hold beyond leading order in the parton density~\cite{Balitsky2,KV}. Secondly, a combined comprehensive analysis of HERA 
and RHIC data is still lacking though there have been first attempts in this direction~\cite{KugeratskiGN1}. 

\section{The future of small $x$ physics at hadron colliders and DIS}

\begin{figure}[htbp]
\begin{center}
\includegraphics[width=0.4\linewidth]{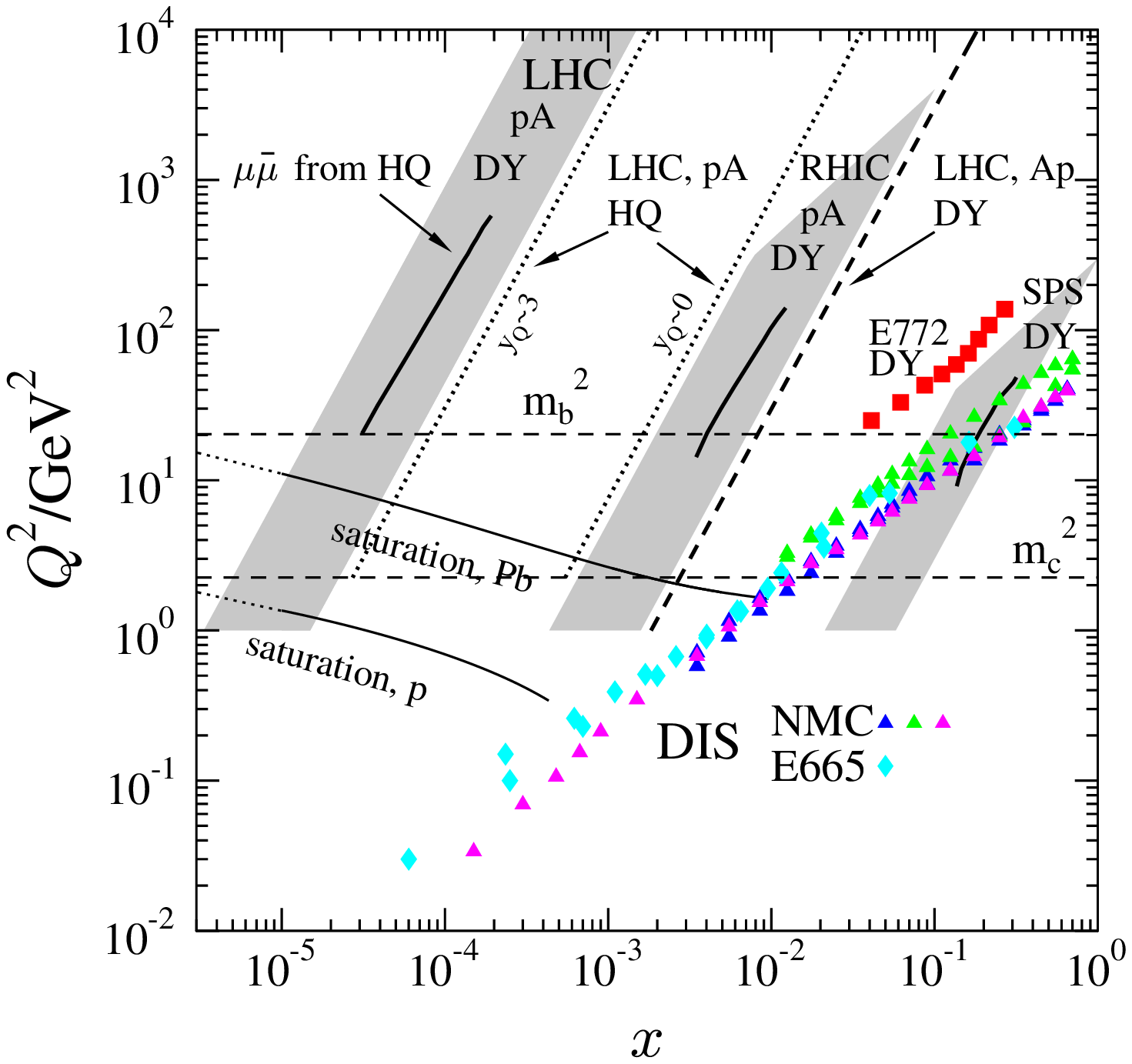}
\includegraphics[width=0.4\linewidth]{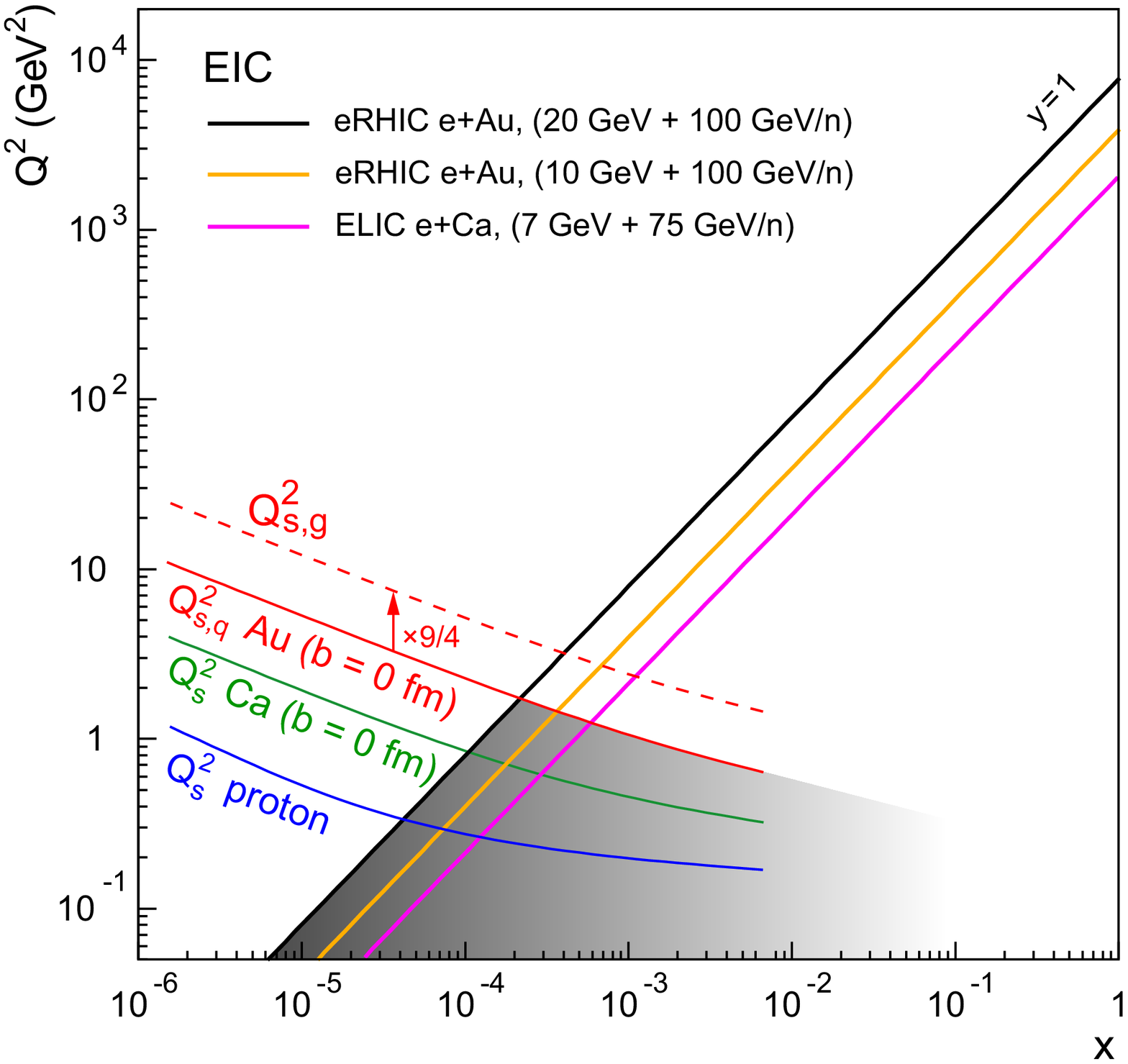}
\end{center}
\caption{Left: Kinematic $x$-$Q^2$ reach of different final states at the LHC compared to other experiments with 
nuclei . From ~\cite{D'Enterria}. Right: The saturation 
scale in the proton, calcium and gold in the kinematic acceptance of the EIC.}
\label{fig:LHC-EIC}
\end{figure}

The LHC is the ultimate small $x$ machine in terms of reach in $x$ for large $Q^2$. A plot from 
Ref.~\cite{D'Enterria} illustrating this reach is shown in Fig.~\ref{fig:LHC-EIC} (left). For a recent review of the small $x$ opportunities at the LHC, see Ref.~\cite{WeissFS}. The LHC will provide further, more extensive tests of the hints for the CGC seen at RHIC. At very high energies, a novel ``diffusive scaling" regime has been proposed, which incorporates the physics of Pomeron loops~\cite{HIMST}. Recent developments were reviewed at DIS06 by Iancu and at DIS07 by Shoshi~\cite{Edmond};  possible signatures at the LHC have been proposed~\cite{IMS}. However, very recent computations including running coupling effects suggest that this regime is unlikely to be accessed realistic collider energies~\cite{Dumitru-etal}.

The universality of parton distributions is often taken for granted but factorization theorems proving this universality have been proven only for a limited number of inclusive final states. However, as we have discussed, small $x$ is the domain of rich multi-parton correlations. These are more sensitive to more exclusive final states for which universality is not proven~\cite{CollinsQiu}. Therefore, while the LHC will have unprecedented reach in $x$, precision studies of high energy QCD and clean theoretical interpretations of these motivate future DIS projects. Two such projects discussed at this conference 
are the EIC project in the United States~\cite{Surrow} and the LHeC project in Europe~\cite{Newman}. 

As we discussed previously, strong color fields may be more easily accessible in DIS off nuclei relative to the proton due to the ``oomph" factor. In Fig.~\ref{fig:LHC-EIC} (right), we show the saturation scale $\qssa(x)$ overlaid on the $x$-$Q^2$ 
kinematic domain spanned by the EIC. It is interesting that there is a significant kinematic domain where $\qssa > Q^2$, 
including in particular $\qssa > 1$ GeV$^2$. In the weak field regime where $Q^2 \gg \qssa$, we are accustomed to thinking of $\alpha_S\equiv \alpha_S(Q^2)$. In the strong field regime, where $\qssa \gg Q^2$, we likely have instead $\alpha_S\equiv \alpha_S(\qssa)$. 
As suggested by the figure, the EIC (and clearly the LHeC) will cleanly probe the cross-over regime from 
linear to non-linear dynamics in QCD. A particularly striking feature of e+A DIS will be diffractive scattering~\cite{NZFSLLKG,HTR};  it is anticipated that $\sim 30\%$ of the cross-section corresponds to hard diffractive final states. For further discussion of the physics of an Electron Ion collider, see Ref.~\cite{ARRW}. 

\section{From CGC to QGP: how classical fields decay in the exploding Glasma}
\label{sec:Glasma}

The word ``Glasma"  describes the strongly interacting matter in heavy ion collisions from the time when particles are produced in the shattering of two CGCs to the time when a thermalized QGP is formed~\cite{LappiMcLerran}. 
We will discuss here a systematic approach to computing particle production in heavy ion collisions to NLO. This 
approach suggests a deep connection between quantum evolution effects in the nuclear wavefunction and instabilities that 
may be responsible for fast thermalization of the Glasma. 
\begin{figure}[htbp]
\begin{center}
\resizebox*{4cm}{!}{\includegraphics{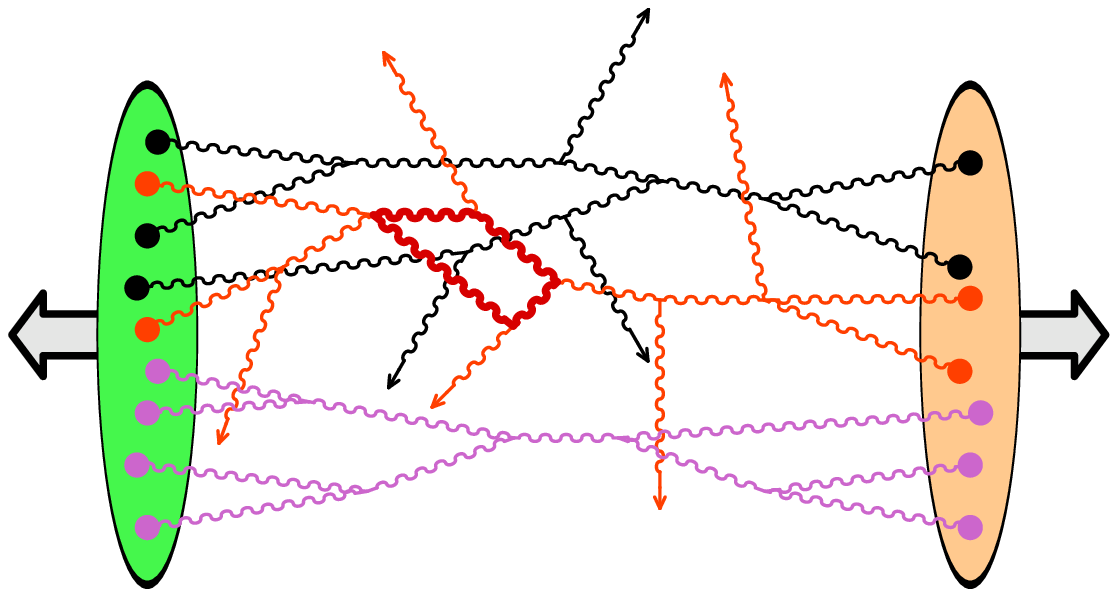}}
\end{center}
\caption{\label{fig:cartoon_1}Cartoon of gluon production in
the collision of two sheets of Colored Glass. The dots denote large $x$ color sources.}
\end{figure}  

A cartoon of multi-particle production in a heavy ion collision is shown in Fig.~\ref{fig:cartoon_1}.  
The probability of producing $n$ particles, in field theories (such as the CGC) with strong external sources can 
be expressed as~\cite{Gelis}
\begin{equation}
P_n = \exp\left(-{1\over g^2}\sum_r b_r\right)\, \sum_{p=1}^n {1\over p!} \sum_{\alpha_1+\cdots \alpha_p = n} {b_{\alpha_1}\cdots b_{\alpha_p}\over g^{2p}} \, .
\label{eq:MP1}
\end{equation}
where $b_r$ denotes the sum of vacuum-to-vacuum graphs with r cuts. This formula has remarkable features: 
a) $P_n$ is non-perturbative in $g$ even for $g\ll1$-no simple power expansion in terms of $g$ exists. b) $P_n$, for any $n$, gets contributions from cut tree vacuum graphs-this would not apply 
for field theories in the vacuum. c) Even at tree level, $P_n$ is {\it not} a Poisson distribution which is counter to presumptions that  classical field theories have only ``trivial" Poissonian correlations. The simple formula in Eq.~\ref{eq:MP1} 
contains many features of the well known AGK calculus~\cite{AGK} of multi-particle production.
  
Computing the probabilities in Eq.~\ref{eq:MP1} is hopeless even for $g\ll1$. Fortunately, a systematic expansion in powers of g exists for moments of the multiplicity. Both the LO and NLO multiplicity can be represented in terms of solutions of equations of motion with retarded boundary conditions. At leading 
order, these are solutions $A_{\rm cl.}^{a,\mu}$ to the Yang-Mills equations; these equations, with {\it boost invariant} CGC initial conditions, were solved numerically in Ref.~\cite{KV,KNV} to compute the gauge fields at late times. 

A next-to-leading order (NLO) computation is important to understand the renormalization and factorization issues that are fundamental to any quantum field theory. As we shall discuss shortly, it is also important to understand the quantum fluctuations that generate the plasma instabilities which may speed up thermalization. Remarkably, the NLO contributions can be computed by solving the initial value problem of small fluctuation equations of motion with retarded boundary conditions. A similar algorithm has been constructed and implemented to study quark pair production in the classical CGC background field~\cite{GLK}. 

In the Glasma, the classical LO {\it boost invariant} $E$ and $B$ fields are purely longitudinal at $\tau=0$. The corresponding momentum distributions, at $\tau > 1/Q_s$, are very unstable--indeed, they lead to an instability which may be analogous to the well known Weibel instability in electromagnetic plasmas. For a review and relevant references, see Ref.~\cite{StanMike}.  3+1-D numerical simulations demonstrate that small rapidity dependent quantum fluctuations grow exponentially and generate longitudinal pressure~\cite{Paul}. The  initial ``seed for 
the simulations corresponded to ``white noise" Gaussian random fluctuations. The maximally unstable modes of the longitudinal pressure grow as $\exp\left(C\sqrt{\Lambda_s \tau}\right)$ with $C\approx 0.425$; this form of the growth was previously predicted for Weibel instabilities in expanding plasmas~\cite{ALM}. Albeit the solutions of the Yang-Mills equations display similar features to  the HTL studies, a deeper understanding of this connection is elusive. 
 
First  quantum corrections to the classical background field of two nuclei at $\tau=0$ give initial conditions~\cite{FGM} that are quite different from those in Ref.~\cite{Paul}. Simulations are underway to determine whether these initial conditions speed up thermalization. A full treatment of quantum fluctuations requires that we understand how some NLO contributions are absorbed in the evolution of the nuclear wavefunctions with energy while the rest contribute to gluon production. A proof of this high energy ``factorization" is in progress~\cite{GLV}. To fully understand fast thermalization in the presence of instabilities, one also needs a kinetic theory of the Glasma that describes the decay of classical fields into particles. A first step has been made in this direction~\cite{GJV}.


\begin{thebibliography}{99}

\bibitem{Wilczek}F. Wilczek, hep-ph/9907340.

\bibitem{Lattice-review}P. Weisz, {\it these proceedings}.

\bibitem{Vogt}A. Vogt, {\it these proceedings}.

\bibitem{GLRMQ}
L.~V. Gribov, E.~M. Levin and M.~G. Ryskin,
\, Phys. Rept. {\bf 100}, 1 (1983);
A.~H. Mueller and J.-W. Qiu,
\, Nucl. Phys. {\bf B268}, 427 (1986).

\bibitem{CGC}{L.D. McLerran, R. Venugopalan}, Phys. Rev. {\bf D 49}, 2233 (1994); 
{\it ibid.}, {\bf D 49}, 3352 (1994); {\it ibid.} {\bf D 50}, 2225 (1994); 
{Yu.V. Kovchegov}, Phys. Rev. {\bf D 54}, 5463 (1996); {J. Jalilian-Marian, A. Kovner, L.D. McLerran, H. Weigert}, Phys. Rev. {\bf D 55}, 5414 (1997); E.~Iancu, A.~Leonidov and L.~D.~McLerran,
  Nucl.\ Phys.\  A {\bf 692}, 583 (2001).

\bibitem{JIMWLK}{J. Jalilian-Marian, A. Kovner, L.D. McLerran, H. Weigert}, Phys. Rev. {\bf D 55}, 5414 (1997); 
{J. Jalilian-Marian, A. Kovner, A. Leonidov, H. Weigert}, Nucl. Phys. {\bf B 504}, 415 (1997); {\it ibid.},Phys. Rev. {\bf D 59}, 034007 (1999); E.~Iancu, A.~Leonidov and L.~D.~McLerran,
  Nucl.\ Phys.\  A {\bf 692}, 583 (2001); E.~Ferreiro, E.~Iancu, A.~Leonidov and L.~McLerran,
  Nucl.\ Phys.\  A {\bf 703}, 489 (2002).

\bibitem{BK}{I. Balitsky}, Nucl. Phys. {\bf B 463}, 99 (1996); 
{Yu.V. Kovchegov}, Phys. Rev. {\bf D 61}, 074018 (2000).

\bibitem{GolecSK}A.~M.~Stasto, K.~Golec-Biernat and J.~Kwiecinski,
  Phys.\ Rev.\ Lett.\  {\bf 86}, 596 (2001).

\bibitem{MarquetSchoeffel}C.~Marquet and L.~Schoeffel,
  Phys.\ Lett.\  B {\bf 639}, 471 (2006).

\bibitem{Gelis-etal}F.~Gelis, R.~Peschanski, G.~Soyez and L.~Schoeffel,
  Phys.\ Lett.\  B {\bf 647}, 376 (2007).

\bibitem{MunierPeschanski}S.~Munier and R.~Peschanski,
  Phys.\ Rev.\ Lett.\  {\bf 91}, 232001 (2003); Phys.\ Rev.\  D {\bf 69}, 034008 (2004); {\it ibid.},{\bf 70}, 077503 (2004).

\bibitem{AlbaceteKovchegov}J.~L.~Albacete and Y.~V.~Kovchegov,
  Phys.\ Rev.\  D {\bf 75}, 125021 (2007).

\bibitem{AvsarGustafson}E.~Avsar and G.~Gustafson,
  JHEP {\bf 0704}, 067 (2007).

\bibitem{Cyrille}C.~Marquet, R.~Peschanski and G.~Soyez,
  Phys.\ Lett.\  B {\bf 628}, 239 (2005).

\bibitem{Mueller:1989st}
A.~H. Mueller,
\, Nucl. Phys. {\bf B335}, 115 (1990);
N.~N. Nikolaev and B.~G. Zakharov,
\, Phys. Lett. {\bf B260}, 414 (1991);
\, Z. Phys. {\bf C49}, 607 (1991).
\, Z. Phys. {\bf C53}, 331 (1992).

\bibitem{Golec-Biernat:1998js}
K.~Golec-Biernat and M.~Wusthoff,
\, Phys. Rev. {\bf D59}, 014017 (1999); 
\, Phys. Rev. {\bf D60}, 114023 (1999).

\bibitem{Venugopalan:1999wu}
L.~D. McLerran and R.~Venugopalan,
\, Phys. Rev. {\bf D59}, 094002 (1999); R.~Venugopalan,
\, Acta Phys. Polon. {\bf B30}, 3731 (1999).

\bibitem{Kowalski:2003hm}
H.~Kowalski and D.~Teaney,
\, Phys. Rev. {\bf D68}, 114005 (2003).

\bibitem{Bartels:2002cj}
J.~Bartels, K.~Golec-Biernat and H.~Kowalski,
\, Phys. Rev. {\bf D66}, 014001 (2002).

\bibitem{Iancu:2003ge}
E.~Iancu, K.~Itakura and S.~Munier,
\, Phys. Lett. {\bf B590}, 199 (2004).

\bibitem{Kowalski:2006hc}
H.~Kowalski, L.~Motyka and G.~Watt,
\, Phys. Rev. {\bf D74}, 074016 (2006).

\bibitem{Forshaw:2006np}
J.~R. Forshaw, R.~Sandapen and G.~Shaw,
\, JHEP {\bf 11}, 025 (2006).

\bibitem{Thorne}R.~S.~Thorne,
  Phys.\ Rev.\  D {\bf 71}, 054024 (2005); C.~Ewerz and O.~Nachtmann,
  Annals Phys.\  {\bf 322}, 1635 (2007).

\bibitem{Golec-Sapeta}K.~Golec-Biernat and S.~Sapeta,
  Phys.\ Rev.\  D {\bf 74}, 054032 (2006).

\bibitem{Soyez-07}G.~Soyez,
  arXiv:0705.3672 [hep-ph]; C.~Marquet,
  arXiv:0706.2682 [hep-ph].

\bibitem{FadinBWK}V.~S.~Fadin, R.~Fiore, A.~V.~Grabovsky and A.~Papa,
  arXiv:0705.1885 [hep-ph]; V.~S.~Fadin, R.~Fiore and A.~Papa,
  Phys.\ Lett.\  B {\bf 647}, 179 (2007); Nucl.\ Phys.\  B {\bf 769}, 108 (2007); I.~Balitsky,
  Phys.\ Rev.\  D {\bf 75}, 014001 (2007); Y.~V.~Kovchegov and H.~Weigert,
  Nucl.\ Phys.\  A {\bf 789}, 260 (2007); {\it ibid.},  Nucl.\ Phys.\  A {\bf 784}, 188 (2007); E.~Gardi, J.~Kuokkanen, K.~Rummukainen and H.~Weigert,
  Nucl.\ Phys.\  A {\bf 784}, 282 (2007).

\bibitem{DonnachieLandshoff}A.~Donnachie and P.~V.~Landshoff,
  Phys.\ Lett.\  B {\bf 437}, 408 (1998).

\bibitem{Freund:2002ux}
A.~Freund, K.~Rummukainen, H.~Weigert and A.~Schafer,
\, Phys. Rev. Lett. {\bf 90}, 222002 (2003).

\bibitem{HTR}H.~Kowalski, T.~Lappi and R.~Venugopalan,
  arXiv:0705.3047 [hep-ph].

\bibitem{Mueller:2003bz}
A.~H. Mueller,
\, Nucl. Phys. {\bf A724}, 223 (2003).

\bibitem{Armesto:2004ud}
N.~Armesto, C.~A. Salgado and U.~A. Wiedemann,
\, Phys. Rev. Lett. {\bf 94}, 022002 (2005).

\bibitem{KovnerMW}{A. Kovner, L.D. McLerran, H. Weigert}, Phys. Rev. {\bf D 52}, 3809 (1995); 
 {\bf D 52}, 6231 (1995); 
{Yu.V. Kovchegov, D.H. Rischke}, Phys. Rev. {\bf C 56}, 1084 (1997).

\bibitem{KV}{A. Krasnitz, R. Venugopalan}, Nucl. Phys. {\bf B 557}, 237 (1999); 
 Phys. Rev. Lett. {\bf 84}, 4309 (2000);  {\bf 86}, 1717 (2001).
 
\bibitem{KN}D. Kharzeev, M. Nardi, Phys. Lett. {\bf B 507}, 121 (2001).
 
\bibitem{KL}D. Kharzeev, E. Levin, Phys. Lett. {\bf B 523}, 79 (2001).

\bibitem{KNV}A.~Krasnitz, Y.~Nara and R.~Venugopalan,
  Phys.\ Rev.\ Lett.\  {\bf 87}, 192302 (2001); Nucl.\ Phys.\ A {\bf 717}, 268 (2003); {\it ibid.}, {\bf 727}, 427 (2003); T.~Lappi,
  Phys.\ Rev.\ C {\bf 67}, 054903 (2003); Phys.\ Lett.\ B {\bf 643}, 11 (2006); K.~Fukushima,
  arXiv:0704.3625 [hep-ph].
 
\bibitem{LappiV}T.~Lappi and R.~Venugopalan,
  Phys.\ Rev.\  C {\bf 74}, 054905 (2006).

\bibitem{Hirano-etal}T. Hirano, U.W. Heinz, D. Kharzeev, R. Lacey, Y. Nara, Phys. Lett. {\bf B 636}, 299 (2006); 
H.-J. Drescher, A. Dumitru, A. Hayashigaki, Y. Nara, Phys. Rev. {\bf C 74}, 044905 (2006). 

\bibitem{KovchegovMueller}Y.~V.~Kovchegov and A.~H.~Mueller,
  Nucl.\ Phys.\  B {\bf 529}, 451 (1998); A.~Dumitru and L.~D.~McLerran,
  Nucl.\ Phys.\  A {\bf 700}, 492 (2002); J.~P.~Blaizot, F.~Gelis and R.~Venugopalan,
  Nucl.\ Phys.\  A {\bf 743}, 13 (2004); F.~Gelis and Y.~Mehtar-Tani,
  Phys.\ Rev.\  D {\bf 73}, 034019 (2006).
  
\bibitem{Jalilian}J.~Jalilian-Marian, Phys. Rev. {\bf C 70}, 027902 (2004).
  
\bibitem{FAR}F.~Gelis, A.~M.~Stasto and R.~Venugopalan,
  Eur.\ Phys.\ J.\  C {\bf 48}, 489 (2006).
  
\bibitem{EKL}E.~Iancu, K.~Itakura and L.~McLerran,
  Nucl.\ Phys.\  A {\bf 721}, 293 (2003).

\bibitem{KharzeevLM}D.~Kharzeev, E.~Levin and L.~McLerran,
  Phys.\ Lett.\  B {\bf 561}, 93 (2003).

\bibitem{KKT}D. Kharzeev, Yu. Kovchegov, K. Tuchin, 
Phys. Rev. {\bf D 68}, 094013 (2003). 

\bibitem{DumitruHJ}A.~Dumitru, A.~Hayashigaki and J.~Jalilian-Marian,
  Nucl.\ Phys.\  A {\bf 770}, 57 (2006).

\bibitem{BoerDH}D.~Boer, A.~Dumitru and A.~Hayashigaki,
  Phys.\ Rev.\  D {\bf 74}, 074018 (2006).

\bibitem{YJ2}{J. Jalilian-Marian, Y. Kovchegov}, Prog. Part. Nucl. Phys. {\bf 56}, 104 (2006).

\bibitem{BlaizotGV2}J.P. Blaizot, F. Gelis, R. Venugopalan, 
Nucl. Phys. {\bf A 743}, 57 (2004); H.~Fujii, F.~Gelis, R.~Venugopalan, 
Phys. Rev. Lett.  {\bf 95}, 162002 (2005); Nucl.\ Phys.\  A {\bf 780}, 146 (2006); K.~Tuchin,
  Phys.\ Lett.\  B {\bf 593}, 66 (2004); {N.N. Nikolaev, W. Schafer}, Phys. Rev. {\bf D 71}, 014023 (2005); 
{N.N. Nikolaev, W. Schafer, B.G. Zakharov}, Phys. Rev. Lett. {\bf 95}, 221803 (2005). 

\bibitem{GelisV1}F.~Gelis, R.~Venugopalan, Phys.\ Rev. {\bf D 69}, 014019 (2004).

\bibitem{Balitsky2}I.~Balitsky,
  Phys.\ Rev.\  D {\bf 72}, 074027 (2005); J.~Jalilian-Marian and Y.~V.~Kovchegov,
  Phys.\ Rev.\  D {\bf 70}, 114017 (2004)
  [Erratum-ibid.\  D {\bf 71}, 079901 (2005)].

\bibitem{KugeratskiGN1}V.~P. Goncalves, M.~S. Kugeratski, M.~V.~T. Machado and F.~S. Navarra,
\, Phys. Lett. {\bf B643}, 273 (2006).

\bibitem{D'Enterria}D.~d'Enterria,
  Eur.\ Phys.\ J.\  A {\bf 31}, 816 (2007).

\bibitem{WeissFS}L.~Frankfurt, M.~Strikman and C.~Weiss,
  Ann.\ Rev.\ Nucl.\ Part.\ Sci.\  {\bf 55}, 403 (2005).

\bibitem{HIMST}Y.~Hatta, E.~Iancu, C.~Marquet, G.~Soyez and D.~N.~Triantafyllopoulos,
  Nucl.\ Phys.\  A {\bf 773}, 95 (2006).

\bibitem{Edmond}E.~Iancu,
  arXiv:hep-ph/0608086; A.~I.~Shoshi,
  arXiv:0706.1866 [hep-ph].

\bibitem{IMS}E.~Iancu, C.~Marquet and G.~Soyez,
  Nucl.\ Phys.\  A {\bf 780}, 52 (2006); M.~Kozlov, A.~I.~Shoshi and B.~W.~Xiao,
  arXiv:hep-ph/0612053.

\bibitem{Dumitru-etal}A.~Dumitru, E.~Iancu, L.~Portugal, G.~Soyez and D.~N.~Triantafyllopoulos,
  arXiv:0706.2540 [hep-ph].

\bibitem{CollinsQiu}J.~Collins and J.~W.~Qiu,
  Phys.\ Rev.\  D {\bf 75}, 114014 (2007).

\bibitem{Surrow}B. Surrow, {\it these proceedings}.

\bibitem{Newman}P. Newman, {\it these proceedings}.

\bibitem{NZFSLLKG}N.~N.~Nikolaev, B.~G.~Zakharov and V.~R.~Zoller,
  Z.\ Phys.\  A {\bf 351}, 435 (1995); L.~Frankfurt, V.~Guzey and M.~Strikman,
  Phys.\ Lett.\  B {\bf 586}, 41 (2004); E.~Levin and M.~Lublinsky,
  Nucl.\ Phys.\  A {\bf 712}, 95 (2002); M.~S.~Kugeratski, V.~P.~Goncalves and F.~S.~Navarra,
  Eur.\ Phys.\ J.\  C {\bf 46}, 413 (2006).

\bibitem{ARRW}A.~Deshpande, R.~Milner, R.~Venugopalan and W.~Vogelsang,
  Ann.\ Rev.\ Nucl.\ Part.\ Sci.\  {\bf 55}, 165 (2005).

\bibitem{Gelis}F.~Gelis and R.~Venugopalan,
  Nucl.\ Phys.\ A {\bf 776}, 135 (2006); {\it ibid.} {\bf 779}, 177 (2006); arXiv:hep-ph/0611157; F. Gelis, arXiv:hep-ph/0701225.
  
\bibitem{LappiMcLerran}T.~Lappi and L.~McLerran,
  Nucl.\ Phys.\ A {\bf 772}, 200 (2006); D.~Kharzeev, A.~Krasnitz and R.~Venugopalan,
  Phys.\ Lett.\ B {\bf 545}, 298 (2002).
  
\bibitem{GLK}F.~Gelis, K.~Kajantie and T.~Lappi,
  Phys.\ Rev.\ C {\bf 71}, 024904 (2005); Phys.\ Rev.\ Lett.\  {\bf 96}, 032304 (2006).
  
\bibitem{AGK}{V.A. Abramovsky, V.N. Gribov, O.V. Kancheli}, Sov. J. Nucl. Phys. {\bf 18},
  308 (1974).  
  
\bibitem{StanMike}S.~Mrowczynski and M.~H.~Thoma,
  arXiv:nucl-th/0701002; M.~Strickland,
  arXiv:hep-ph/0701238.

\bibitem{Paul}P.~Romatschke and R.~Venugopalan,
  Phys.\ Rev.\ Lett.\  {\bf 96}, 062302 (2006); Eur.\ Phys.\ J.\ A {\bf 29}, 71 (2006); Phys.\ Rev.\ D {\bf 74}, 045011 (2006).

\bibitem{ALM}P.~Arnold, J.~Lenaghan and G.~D.~Moore,
  JHEP {\bf 0308}, 002 (2003).

\bibitem{FGM}K.~Fukushima, F.~Gelis and L.~McLerran,
  Nucl.\ Phys.\  A {\bf 786}, 107 (2007).

\bibitem{GLV}F. Gelis, T. Lappi and R. Venugopalan, in preparation.

\bibitem{GJV}F.~Gelis, S.~Jeon and R.~Venugopalan,
  arXiv:0706.3775 [hep-ph].  

  
    
  
  
  
  
  
 
 
 


  









\end{thebibliography}
\end{document}